\documentstyle[preprint,aps,epsfig]{revtex}

\begin{document}
\draft

\title{Prospecting for the superfluid transition in electron--hole coupled quantum
wells using Coulomb drag}
\author{Ben Yu-Kuang Hu}
\address{Department of Physics, University of Akron, 
Akron, OH 44325-4001}

\date{\today}
\maketitle

\begin{abstract}
I investigate the possibility of using Coulomb drag to detect a precursor of the 
predicted (but as yet not definitively observed) superfluid transition in electron--hole 
coupled quantum wells.
The drag transresisitivity $\rho_{21}$ is shown to be significantly enhanced above
the transition temperature $T_c$ 
and to diverge logarithmically as $T\rightarrow T_c^+$, 
due to electron--hole pairing fluctuations which are somewhat analogous to the Maki-Thompson
contribution to conductivity in metals above the superconducting $T_c$.
The enhancement in $\rho_{21}$ is 
estimated to be detectable at temperatures much higher than $T_c$.  
\end{abstract}

Under suitable conditions, an attractive interaction between fermions  
causes a pairing instability that leads to phase transition, such as  
superconductivity, 
where phonons mediate an effective attractive interaction between electrons.   
Since the direct Coulomb interaction is in general stronger than effective 
phonon-mediated interactions, a coupled electron--hole gas might be expected to 
readily undergo an equivalent phase transition.
This possibility was theoretically investigated in the sixties,
and it was predicted that electron--hole systems in bulk systems would 
form an condensate at low enough temperatures\cite{jero67,russkies}. 
Despite a formal similarity between electron--hole condensate and the 
BCS superconducting states, the former was shown to be in fact an insulator. 
Subsequently, it was proposed\cite{russkies1} 
that {\em separated} electron--hole systems could form a superfluid state, 
due to the suppresion of 
transitions that fix the phase of the order parameter thereby producing an insulator.
The separation of the electron and hole components
in semiconductor systems also has an important experimental advantage;
it inhibits the recombination of the electrons and holes which would otherwise
rapidly deplete the system.

The advent of fabrication techniques such as molecular beam epitaxy has made it
possible to produce high quality electron--hole double well systems and to test these 
predictions.  Experimental searches for double layer electron--hole
condensation have generally used optical techniques.  While there are 
reports of optical signatures\cite{exp} hinting at a transition, there has been no 
definitive observation of a superfluid condensation.
Recently, Vignale and MacDonald\cite{vign96} proposed the superfluid transition
could be definitively identified through a completely different route: using Coulomb drag.  

In Coulomb drag experiments, each well in the coupled
well system is individually electrically contacted,  
making it possible to isolate a driving voltage in just one of the two wells.   
When an electrical current is driven through one well, the interlayer interactions
produce a measurable drag signal in the other.  The measured quantity is
typically the transresistivity $\rho_{21}=E_2/J_1$, where $J_1$ is the current
density in the driving layer and $E_2$ is the electric field response in the
other layer when it in an open circuit.  
There has been considerable experimental\cite{dragexp} and
theoretical\cite{dragtheo} activity in drag in the past decade.

For most typical experimental parameters,
the interlayer interaction is a weak perturbation, in
which case drag is adequately described
within the Born approximation; {\em i.e.}, the
carriers in the adjacent wells are uncorrelated, 
and carriers in one well occasionally collide with carriers
in the other.  Typically, $\rho_{21}$ is much smaller
than the resistivities of the individual layers, $\rho_{11}$ and $\rho_{22}$.
Furthermore, in this weak-coupling regime the Born approximation unambiguously predicts 
$\rho_{21}\rightarrow 0$ as the temperature $T$ goes to zero, because
the scattering phase space for the quasiparticles vanishes at zero temperature.
What Vignale and MacDonald\cite{vign96} showed is that when electron--hole condensation 
occurs, $\rho_{21}$ becomes {\em comparable} in magnitude to $\rho_{11}$ and $\rho_{22}$,
and {\em diverges} when $T\rightarrow 0$.
As yet, this has not been experimentally observed. 

Even when the electrons and holes are not in a condensed state, interlayer correlations 
can affect the transresisitivity\cite{swie95}.
In the closely related case of drag of composite fermions (electrons bound to two
magnetic flux quanta), 
the build-up of interlayer correlations produces unique signatures in 
the transresistivity\cite{uddi98} which differ markedly from the weak-coupling result
$\rho_{21}(T\rightarrow 0)=0$.
Deviations from the weak-coupling result in composite fermion drag
at low temperatures have in fact been reported\cite{lill98}.
Analogously, deviations from the weak-couling form of $\rho_{21}(T)$ in electron--hole
systems would indicate the
presence of interlayer correlations which foreshadow an electron--hole condensation;
thus Coulomb drag can be used to ``prospect" for situations where electron--hole
condensation could occur.
 
In this paper, I evaluate the transresistivity in the normal state
of an electron--hole double quantum well system, going beyond the
Born approximation and taking account correlations
between the two wells.  
I use the Kubo formalism to calculate 
$\sigma_{21} = J_1/E_2$, from which the transresistivity is given by
$\rho_{21} \approx -\sigma_{21}(\sigma_{11}\sigma_{22})^{-1}$.
The Hamiltonian I use to describe this system is
\begin{eqnarray}
H &=& \sum_{i=1,2}\sum_{{\bf k}_i} \epsilon_{i}({\bf k}_i)\,
\hat{c}_i^\dagger({\bf k}_i)\,\hat{c}_i({\bf k}_i) 
+ \sum_{\bf q}
\hat\rho_1({\bf q})\,\hat\rho_2({\bf -q})\, V({\bf q}) + \hat{H}_{\rm imp}.
\end{eqnarray}
Here $\hat{c}_i^{(\dagger)}$ are the particle field operators of layer 
$i = 1,2$, (the drive and
drag layers, respectively), $\hat\rho_i$ is the particle
density operator and 
$\hat{H}_{\rm imp}$ is the contribution from static impurities, which are assumed
to be uncorrelated\cite{gorn99}.  
``Particle" refers to either electron or hole.  Within linear response, 
it does not matter which layer is driven.
The static transconductivity $\sigma_{21} = -e_1 e_2
\lim_{\omega\rightarrow 0}\;{\rm Im}[\Pi(\omega)/\omega]$ 
where $e_i$ are the charges in layer $i$ and
$\Pi(i\Omega_n) = {\cal A}^{-1}\int_0^{T^{-1}} d\tau\ 
e^{i\Omega_n\tau}\, \langle\hat{j}_1(i\tau)\hat{j}_2(0)\rangle$
is the current--current correlation function (${\cal A}$ = sample
area, $\Omega_n$ = Matsubara boson frequency).  In this paper, $\hbar, k_B = 1$.

To evaluate $\sigma_{21}$, the standard diagrammatic techniques\cite{flen95,kame95} are used.
The terms I concentrate on are related to the Maki-Thompson terms used
to study effects of superconducting fluctuations in the conductivity of metals
above $T_c$, and the Feynman diagrams are shown in Figs.\ 1(a,b). 
Physically, these terms describe ``the effect of ephemeral Cooper pairs on the 
conductivity"\cite{crav73}.
Composite fermion drag\cite{uddi98,lill98} differs from electron--hole drag in that
the composite fermions have relatively large masses and the phase-breaking processes are 
dominant\cite{bones}, which favor the Asmalazov-Larkin terms\cite{crav73,kell72}.
It is assumed in this paper that the samples are clean and phase-breaking
processes are weak.  The correlations are described by ${\cal T}$-matrices which are 
given by the Bethe-Salpeter equation, shown in Figs.\ 1(c,d).
${\cal T}_{pp}$ and ${\cal T}_{pa}$ correspond to the particle--particle (pp) and 
and particle-antiparticle (pa) channel, respectively.
(The term ``antiparticle" rather than the standard ``hole" is used to avoid
the dual use of the word ``hole.") 

If the interlayer interaction is assumed to be static, then the
frequency dependence of ${\cal T}_{pp}$
(${\cal T}_{pa}$) comes only from the sum (difference) of the energies 
in the vertices. 
Then, from standard diagrammatic rules\cite{maha90a},
\begin{mathletters}
\begin{equation}
\Pi(i\Omega_n)
= - \frac{4}{{\cal A}^2}\sum_{\bf k_1 k_2}
{v}_{1,x}({\bf k}_1)\,{v}_{2,x}({\bf k}_2)\;
F({\bf k}_1,{\bf k}_2;i\Omega_n),\label{eq:1}
\end{equation}
\begin{eqnarray}
F({\bf k}_1,{\bf k}_2;i\Omega_n)&&= T^2 \sum_{ik_{i,n}}
G_1({\bf k}_1,ik_{1,n}+i\Omega_n) G_1({\bf k}_2,ik_{i,n})
\sum_{ik_{2,n}}
G_2({\bf k}_2,ik_{2,n}+i\Omega_n) G_2({\bf k}_2,ik_{2,n})
\nonumber\\
&&
\Bigl[
\langle {\bf k}_{pp} | {\cal T}_{pp}\left({\bf P}_{pp};ik_{1,n}+ik_{2,n}+i\Omega_n \right)
| {\bf k}_{pp}\rangle
+
\langle {\bf k}_{pa} | {\cal T}_{pa} \left({\bf P}_{pa};ik_{1,n}-ik_{2,n}\right) |
{\bf k}_{pa} \rangle \Bigr].
\label{eq:2}
\end{eqnarray}
\end{mathletters}
Here, $G_i$ is the particle Green function in layer $i$, and 
$k_{i,n}$ are Matsubara fermion frequencies.
The particle velocities are given by ${\bf v}_i ={\bf k}_i/m_i$.
For convenience, the momentum dependences of the ${\cal T}$-matrices
are parameterized by the center-of-mass coordinates, 
${\bf P}_{pp\atop pa} = {\bf k}_1 \pm {\bf k_2}$ and 
${\bf k}_{pp\atop pa} = x_2{\bf k}_1 \mp x_1{\bf k}_2$,
where $x_i = m_i/(m_1 + m_2)$.

To obtain the static transresisitivty, the analytic continuation
$i\Omega_n \rightarrow \omega+i0^+$, followed by the limit 
$\omega\rightarrow 0$, must be taken.  To simplify the
expressions, I assume that both wells are weakly 
disordered (justified by the excellent quality of the two-dimensional
electron and hole gases produced by molecular beam epitaxy), and 
hence only terms to lowest order in $\tau_i^{-1}$, the inverse particle lifetime,
are kept\cite{simplicity}.  One obtains after some algebra
\begin{eqnarray}
\lim_{\omega\rightarrow 0}{\rm Im}&&
[F_{pp\atop pa}({\bf k}_1,{\bf k}_2;\omega+i0^+)]
\simeq \frac{\omega}{2T}\int_{-\infty}^\infty\frac{d\omega_1'}{2\pi}
\int_{-\infty}^\infty\frac{d\omega_2'}{2\pi}\
A_1^2({\bf k}_1,\omega_1')\; 
A_2^2({\bf k}_2,\omega_2')\nonumber\\
&&
\biggl\{n_F(\omega_1')\; 
n_F(\omega_2')\; n_B(-\omega_1' - \omega_2') 
\langle {\bf k}_{pp}|{\rm Im}[T_{pp}({\bf P};\omega_1'+\omega_2'+i0^+)
|{\bf k}_{pp}\rangle\nonumber\\
&& + n_F(\omega_1')\; n_F(-\omega_2')\; 
n_B(-\omega_1' + \omega_2')\;
\langle {\bf k}_{pa}| {\rm Im}[T_{pa}({\bf P};\omega_1'-\omega_2'+i0^+)] 
|{\bf k}_{pa}\rangle.\biggr\}
\label{eq:Im[F]}
\end{eqnarray}
where $A_i({\bf k},\omega)$ is the spectral function of the particle in
layer $i$ and $n_B$ and $n_F$ are the bose and fermi functions,
respectively.  In the weak disorder limit, 
one can approximate\cite{maha90b}
$A^2_i({\bf k}_i,\omega) \approx 4\pi\tau_i
\delta(\omega-\xi_{{\bf k}_i})$, where $\xi_{{\bf k}_i}$ is the particle kinetic energy 
relative to the chemical potential in layer $i$.
Substituting this into 
Eq.\ (\ref{eq:Im[F]}) and using the generalized optical 
theorem\cite{kadanoff}
\begin{eqnarray}
\langle {\bf k}|{\rm Im}[T_{pp\atop pa}({\bf P}_{pp\atop pa},\omega+i0^+)&&
|{\bf k}\rangle
=
\int \frac{d{\bf q}}{(2\pi)^2}\ 
\Bigl| \langle {\bf k}| T_{pp\atop pa}({\bf P}_{pp\atop pa},\omega+i0^+) |
{\bf k+q}\rangle \Bigr|^2 \,\times\nonumber\\ 
\lim_{i\omega_n\rightarrow \omega+i0^+}&&\left\{{\rm Im}\Bigl[\sum_{{i\overline\omega}_n}
G_1(x_1 {\bf P} + {\bf k},i\omega_n + i\overline\omega_n)
G_2(x_1 {\bf P} - {\bf k},\mp i\overline\omega_n)\Bigr]\right\}
\end{eqnarray}
yields, from Eqs.\ (\ref{eq:1}), (\ref{eq:2}) and the expression for
$\sigma_{21}$ in terms of $\Pi$,
\begin{eqnarray}
&&\sigma_{21} = -\frac{8 \pi \tau_1 \tau_2 e_1 e_2 }{T}
\int \frac{d{\bf k}_1}{(2\pi)^2}
\int \frac{d{\bf k}_2}{(2\pi)^2}
\int \frac{d{\bf q}}{(2\pi)^2}\ 
v_{1,x}({\bf k}_1)\,v_{2,x}({\bf k}_2)\,\Biggr\{
n_F(\xi_{\bf k_1})\, n_F(\xi_{\bf k_2})\, n_F(-\xi_{\bf k_1 +q})\nonumber\\
&&\ n_F(-\xi_{\bf k_2-q})\,
\Bigl|\langle{\bf k}_{pp} | T_{pp}({\bf P}_{pp},\xi_{\bf k_1}+\xi_{\bf
k_2})
|{\bf k}_{pp} + {\bf q}\rangle\Bigr|^2\;\delta\Bigl(\xi_{\bf k_1}+\xi_{\bf k_2} -
\xi_{\bf k_1+q} - \xi_{\bf k_2-q}\Bigr) - n_F(\xi_{\bf k_1}) n_F(-\xi_{\bf k_2})
\nonumber\\
&&\ 
n_F(-\xi_{\bf k_1+q})\,
n_F(\xi_{\bf k_2+q})\,
|\langle{\bf k}_{pa} | T_{pa}({\bf P}_{pa},\xi_{\bf k_1}-\xi_{\bf k_2})
|{\bf k}_{pa} + {\bf q}\rangle|^2 
\delta\Bigl(\xi_{\bf k_1}- \xi_{\bf k_2} -
\{\xi_{\bf k_1+q} - \xi_{\bf k_2+q}\}\Bigr) \Biggr\}.
\label{eq:sigma_21}
\end{eqnarray}

Eq.\ (\ref{eq:sigma_21}) can to a certain extent be interpreted within the 
framework of the semiclassical Boltzmann equation\onlinecite{jauh93}.
The first terms within the integral of Eq.\ (\ref{eq:sigma_21})
corresponds to the contribution of particle--particle scattering 
from state ${\bf k}_1,{\bf k}_2$
with exchange of momentum ${\bf q}$, when the distribution of
layer 1 is shifted by the driving electric field.  
The second term corresponds to the contribution due to Pauli blocking of
electrons going into state ${\bf k}_1+{\bf q}$, or alternatively,
one can look at it as scattering of antiparticles in the Fermi liquid 
in layer 1 from ${\bf k}_1 + {\bf q}$ with particles in  
layer 2.  Note that with a purely semiclassical argument, it is
impossible to have different scattering rates
$T_{pp}$ for particle--particle and $T_{pa}$ for particle--antiparticle collisions,
as is obtained above from a full many-body calculation. 
In the Born approximation limit, $\langle {\bf k} | T_{pp}({\bf P},\omega+i0^+) 
|{\bf k+q}\rangle= \langle {\bf k} | T_{pa}({\bf P},\omega+i0^+) 
| {\bf k + q}\rangle = V(q,\omega)$, the screened interlayer interaction,
one regains the standard weak-coupling
result\cite{flen95,kame95,jauh93,siva92,zhen93} which typically involves a product of the 
individual layer susceptibilities (or, in the general case,
a closely related function\cite{flen95}).
The above shows that when correlations between layers are included,
the expression for $\sigma_{21}$ {\em cannot} be written in the 
standard weak-coupling form.

When the ${\cal T}$-matrix is regular, 
temperature dependence of the transresistivity is not qualitatively different 
from the weak-coupling case.  
For example, for hard-core fermions, as $T\rightarrow 0$ the ${\cal T}$-matrix remains 
finite and consequently
the transresistivity vanishes, in spite of the presence of interlayer correlations.
This result is consistent with the claim that a nonzero transresistivity at $T=0$ is
possible only with a phase transition\cite{yang99}.  When a phase transition such
as superconductivity or electron--hole condensation takes place, the 
${\cal T}$-matrix acquires a singularity at the (mean-field) transition 
temperature $T_c$\cite{schr64}, below which the ${\cal T}$-matrix approximation is invalid.  
Below, I examine $\sigma_{21}$ for $T > T_c$.

For simplicity, to obtain qualitative features
let us assume a simple local interlayer interaction, 
$V(q) = V_0 < 0$.  
Then the ${\cal T}$-matrices within the Bethe-Salpeter approximation are dependent 
only on the total momentum ${\bf P}$ and energy $\omega$ of the 
incoming particles.   The ${\cal T}_{pp}({\bf P},\omega)$ channel
diverges at $T_c$, $\omega=0$ and ${\bf P}=0$.   
Its explicit form is
\begin{mathletters}
\begin{eqnarray}
{\cal T}_{pp}(P,\omega;T) &=& -\frac{|V_0|}
{1 + |V_0|\,\chi_{pp}(P,\omega;T)},
\label{eq:T_ee}\\
\chi_{pp}(P,\omega;T) &=&
\int \frac{d{\bf k}}{(2\pi)^2}\;
\frac{1 - n_F(\xi_{1,x_1{\bf P + k}}) - n_F(\xi_{2,x_2{\bf P-k}})}
{\omega - E_k - {\cal E}_P + \mu_1 + \mu_2 + i0^+}.
\end{eqnarray}
\end{mathletters}
Here, $E_k = k^2(m_1^{-1}+m_2^{-1})/2$ and ${\cal E}_P =
P^2/2(m_1+m_2)$.

The singularity in the ${\cal T}_{pp}$ 
occurs when $\chi_{pp} = -|V_0|^{-1}.$  This occurs at
the highest temperature when $P=0$, $\omega=0$ and $k_{F,1}
= k_{F,2} \equiv k_F$ (matched fermi surfaces).
Expanding $\chi_{pp}(T)$ about $P=0$, $\omega=0$ and 
$\Phi = [k_{F,2}^2 - k_{F,1}^2]/2(m_1+m_2)$ gives
\begin{equation}
{\cal T}_{pp}(P,\omega;T) = -\frac{\rho_{\rm red}}{
\delta\tilde\chi(T) + \alpha_P\, {\cal E}_P + \alpha_\Phi\,
\Phi^2 - i\alpha_\omega\, \omega}
\label{eq:Tpp}
\end{equation}
where $\rho_{\rm red} = m_1m_2/2\pi(m_1 + m_2)$, $\delta\tilde\chi(T) = 2\log(T/T_c)$,
$\alpha_\omega = \pi/(4k_B T)$, $\alpha_\Phi = 7\zeta(3)/(2 T^2\pi^2)\approx 0.43/T^2$ and 
$\alpha_P = 7\zeta(3){\cal E}_{k_F}/(T^2\pi^2)
\approx 0.85 {\cal E}_{k_F}/T^2$\cite{assump}.
Substituting Eq.\ (\ref{eq:Tpp}) into Eq.\ (\ref{eq:sigma_21}) yields
(assuming $\sigma_{21}^2 \ll \sigma_{11}\sigma_{22}$, and reintroducing $h$)
\begin{mathletters}
\begin{eqnarray}
\rho_{21} &=& -\frac{h}{e_1e_2}\,\frac{T^2}{E_{k_F}^2}\;
\frac{8\pi^2}{7\zeta(3)}\,
{\cal I}\left(x_1,\tilde\Phi,s\right)\label{eq:rho21pole}\\
{\cal I}(x_1,\tilde\Phi,s)
&=& \int_{-\infty}^\infty \frac{dy}{y} 
\frac{\tan^{-1}\left(y/s\right)}{[\cosh(y/2)
+ \cosh(y(1/2-x_1) + \tilde\Phi)]^2}
\end{eqnarray}
\end{mathletters}
where 
$s = 8\pi^{-1} \log\left(\frac{T}{T_c}\right) + 14\zeta(3)\pi^{-3}
\tilde\Phi^2$ and 
$\tilde\Phi \equiv\Phi/T_c$.

{\sl For matched electron and hole densities} ({\em i.e.}, $\Phi=0$),
\begin{equation}
{\cal I}\Bigl(x_1,0,s=8\pi^{-1}\log(T/T_c)\Bigr) \simeq \cases{1/s & \mbox{if $s\gg 1$,}\cr
\pi\ln\left(\frac{1}{s}\right) & \mbox{if $s\ll 1$,}\cr}
\end{equation}
ignoring prefactors which depend weakly on $x_1$ and are of order 1.
Thus, the transresistivity diverges logarithmically as $T\rightarrow T_c^+$.
Furthermore, $\rho_{21}(T\gg T_c) \sim 1/\log(T/T_c)$, and this
relatively slow fall-off with increasing temperature
implies that the electron--hole pairing fluctuation enhancement of $\rho_{21}$
can be seen well above $T_c$.  A comparison of Eq.\ 
(\ref{eq:rho21pole}) with the weak-coupling result\cite{jauh93}
$\rho_{21,{\rm weak}} \simeq 
-h T^2\zeta(3)\pi m_1m_2/(8\hbar^2 e_1e_2 
{k}_{F,1}^3 k_{F,2}^3 q_{\rm TF,1} q_{\rm TF,2} d^4),$
implies that the pairing fluctuation contribution to the transresistivity
is bigger than $\rho_{21,{\rm weak}}$ when
\begin{equation}
\log\left(\frac{T}{T_c}\right)
\lesssim
\frac{16\pi}{7[\zeta(3)]^2}
x_1(1-x_1)k_F^2 q_{\rm TF,1} q_{\rm TF,2} d^4.
\label{eq:compare}
\end{equation}
where $q_{\rm TF,i} = 2m_ie_i^2/\kappa\hbar^2$ ($\kappa =$ dielectric constant)
is the Thomas-Fermi screening length in layer $i$
and $d$ is the center-to-center well separation.
For GaAs parameters, with $k_F = 10^6\,{\rm cm}^{-1}$ and $d = 4\times 10^{-6}\,
{\rm cm}$, the number on the right of Eq.~(\ref{eq:compare}) is approximately
$5\times 10^{3}$.
Clearly, this result should not be interpreted quantitatively, since 
several approximations and assumptions have been used in this calculation.
The local interlayer interaction no doubt leads to a significant overestimate of the
pairing fluctuation contribution, since it
fails to cut off the large momentum transfer contributions, unlike the more
realistic interlayer Coulomb interaction.  Furthermore,
the above calculation does not take into account 
other factors that tend to impede electron--hole condensation.
These include uncorrelated impurity potentials in the electron and hole layers
(negating the the time-reversal argument which makes non-magnetic impurity scattering
irrelevant in superconductivity) and band-structure effects\cite{cont98}. 
Nevertheless, the above result strongly suggests that 
the enhancement can be experimentally detected.

{\sl For umatched electron--hole densities} ({\em i.e.}, $\Phi\ne 0$), the density dependence of the pairing fluctuation 
enhanced $\rho_{21}$ differs from the weak-coupling result.  For $k_{F,1}\approx k_{F,2}$, the weak-coupling transresitivity
goes as $\rho_{21,{\rm weak}}\sim (k_{F,1}k_{F,2})^{-3}$; i.e. there is a monotonic increase
with decrease of either density.  On the other hand, since the nesting of the electron 
and hole fermi surfaces is a necessary condition for electron--hole condensation, 
in the pairing fluctuation
enhanced case the transresistivity peaks at matched densities, $\rho_{21}(\Phi=0) -
\rho_{21}(\Phi) \propto (\Phi/T)^2$. 
It should be noted that a peak in the $\rho_{21}$
at $k_{F,1} = k_{F,2}$ is also the signature of a phonon mediated interaction.  
However, the phonon mediated interaction  falls rapidly ($\sim T^6$) below the 
Bloch-Gr{\"u}neisen
temperature\cite{bons98} which is typically a few degrees K for GaAs, so if
a the peak is observed at low enough
temperatures the phonon mechanism can be ruled out.

To conclude, I have presented a strong-coupling theory of Coulomb drag, within the
${\cal T}$-matrix approximation.  Applying this theory to electron--hole double
quantum well systems,
I find that the pairing fluctuations lead to a large enhancement in the 
transresistivity above the electron--hole condensation temperature $T_c$, and
this effect could be used to identify promising candidates for 
the observation of electron--hole condensation.

This work was initiated at Mikroelektronik Centret, Danmarks Tekniske Universitet, and
partially supported by a University of Akron Summer Research Fellowship.
I gratefully acknowledge useful discussions with Martin Chr.~B{\o}nsager, Karsten Flensberg, 
Antti-Pekka Jauho and John W. Wilkins.

\begin{figure}
\phantom{nothing}
\vspace{1cm}
\unitlength=0.8mm
\linethickness{0.8pt}

\begin{picture}(119,25)(-30,100)

\put(0,120){\circle*{3}}
\put(-1.7,125){1}
\put(0,120){\line(1,1){20}}
\put(20,140){\line(0,-1){40}}
\put(0,120){\line(1,-1){20}}
\linethickness{3.2pt}
\put(20,140){\line(1,0){20}}
\put(20,100){\line(1,0){20}}
\linethickness{0.8pt}
\put(40,140){\line(0,-1){40}}
\put(60,120){\line(-1,1){20}}
\put(40,100){\line(1,1){20}}
\put(58.3,125){2}
\put(60,120){\circle*{3}}
\put(25,118){\LARGE{${\cal T}_{pp}$}}
\put(10,130){\line(-1,0){3}}
\put(10,130){\line(0,-1){3}}
\put(10,110){\line(1,0){3}}
\put(10,110){\line(0,-1){3}}
\put(50,130){\line(1,0){3}}
\put(50,130){\line(0,-1){3}}
\put(50,110){\line(-1,0){3}}
\put(50,110){\line(0,-1){3}}
\put(26,90){(a)}

\put(80,120){\circle*{3}}
\put(78.3,125){1}
\put(80,120){\vector(1,1){20}}
\put(100,140){\line(0,-1){40}}
\put(80,120){\line(1,-1){20}}
\linethickness{3.2pt}
\put(100,140){\line(1,0){20}}
\put(100,100){\line(1,0){20}}
\linethickness{0.8pt}
\put(120,140){\line(0,-1){40}}
\put(120,140){\line(1,-1){20}}
\put(120,100){\line(1,1){20}}
\put(138.3,125){2}
\put(140,120){\circle*{3}}
\put(105,118){\LARGE{${\cal T}_{pa}$}}
\put(90,130){\line(-1,0){3}}
\put(90,130){\line(0,-1){3}}
\put(90,110){\line(1,0){3}}
\put(90,110){\line(0,-1){3}}
\put(130,130){\line(-1,0){3}}
\put(130,130){\line(0,1){3}}
\put(130,110){\line(1,0){3}}
\put(130,110){\line(0,1){3}}
\put(106,90){(b)}
\end{picture}

\vspace{2cm}

\unitlength=0.8mm
\newsavebox{\twoelectron}
\savebox{\twoelectron}(20,30){
\begin{picture}(20,30)
\put(9,0){\line(2,1){5}}
\put(9,0){\line(2,-1){5}}
\put(9,30){\line(2,1){5}}
\put(9,30){\line(2,-1){5}}
\put(0,0){\line(1,0){20}}
\put(0,30){\line(1,0){20}}
\end{picture}}

\newsavebox{\electronhole}
\savebox{\electronhole}(20,30){
\begin{picture}(20,30)
\put(0,0){\line(1,0){20}}
\put(9,0){\line(2,1){5}}
\put(9,0){\line(2,-1){5}}
\put(0,30){\line(1,0){20}}
\put(11,30){\line(-2,1){5}}
\put(11,30){\line(-2,-1){5}}
\end{picture}}

\newsavebox{\pptmatrix}
\savebox{\pptmatrix}(20,30){
\begin{picture}(20,30)
\put(0,0){\line(1,0){20}}
\put(0,30){\line(1,0){20}}
\linethickness{3.2pt}
\put(0,0){\line(0,1){30}}
\put(20,0){\line(0,1){30}}
\linethickness{0.8pt}
\put(6,13){\Large${\cal T}_{pp}$}
\end{picture}}

\newsavebox{\patmatrix}
\savebox{\patmatrix}(20,30){
\begin{picture}(20,30)
\put(0,0){\line(1,0){20}}
\put(0,30){\line(1,0){20}}
\linethickness{3.2pt}
\put(0,0){\line(0,1){30}}
\put(20,0){\line(0,1){30}}
\linethickness{0.8pt}
\put(6,13){\Large${\cal T}_{pa}$}
\end{picture}}

\begin{picture}(200,30)(10,0)
\put(-10,13){(c)}
\put(0,0){\usebox{\twoelectron}}
\put(20,0){\usebox{\pptmatrix}}
\put(40,0){\usebox{\twoelectron}}

\put(65,14){\Large{=}}

\put(75,0){\usebox{\twoelectron}}
\thicklines
\put(95,0){\dashbox(0,30){}}
\thinlines
\put(96,14){\Large$V_{0}$}
\put(95,0){\usebox{\twoelectron}}

\put(120,14){\Large{+}}

\put(130,0){\usebox{\twoelectron}}
\put(150,0){\usebox{\pptmatrix}}
\put(170,0){\usebox{\twoelectron}}
\thicklines
\put(190,0){\dashbox(0,30){}}
\thinlines
\put(191,14){\Large$V_{0}$}
\put(190,0){\usebox{\twoelectron}}
\end{picture}

\begin{picture}(200,30)(10,20)
\put(-10,13){(d)}
\put(0,0){\usebox{\electronhole}}
\put(20,0){\usebox{\patmatrix}}
\put(40,0){\usebox{\electronhole}}

\put(65,14){\Large{=}}

\put(75,0){\usebox{\electronhole}}
\thicklines
\put(95,0){\dashbox(0,30){}}
\thinlines
\put(96,14){\Large$V_{0}$}
\put(95,0){\usebox{\electronhole}}

\put(120,14){\Large{+}}

\put(130,0){\usebox{\electronhole}}
\put(150,0){\usebox{\patmatrix}}
\put(170,0){\usebox{\electronhole}}
\thicklines
\put(190,0){\dashbox(0,30){}}
\put(191,14){\Large$V_{0}$}
\thinlines
\put(190,0){\usebox{\electronhole}}
\end{picture}

\vspace{3cm}
\caption{Feynman diagrams used in calculating the transconductivity.
Figs.\ (a) and (b) show the particle--particle and particle--antiparticle
channel contributions to $\sigma_{21}$.  The black dots are the current
vertices, the numbers above them denote the layer indices, and the arrows are
the particle Green functions.  The ${\cal T}_{pp}$ and ${\cal T}_{pa}$
are the particle--particle and particle--antiparticle ${\cal T}$-matrices 
which are calculated using 
the Bethe-Salpeter equation, shown diagramatically in Figs.\ (c) and (d), 
respectively.}
\label{fig_schem}
\end{figure}


\begin{thebibliography}{10}

\bibitem{jero67} D.~J\'erome, T.~M.~Rice, and W.~Kohn, Phys.\ Rev.\ B {\bf 158},
462 (1967).

\bibitem{russkies} L.~V.~Keldysh and Y.~V.~Kopaev, Fiz.\ Tverd.\ Tela {\bf 6},
2791 (1964) [Sov.\ Phys.\ Solid state {\bf 6}, 2219 (1965)]; A.~N.~Kozlov and 
L.~A.~Makzimov, Zh.\ eksp.\ Teor.\ Fiz.\ {\bf 48}, 1184 (1965) [Sov.\ Phys.\
JETP {\bf 21}, 790 (1965)]; L.~V.~Keldysh and A.~N.~Kozlov, Zh.\ Eksp.\ Teor.\
Fiz.\ {\bf 54}, 978 (1968) [Sov.\ Phys.\ JEPT {\bf 27}, 521 (1968)].

\bibitem{russkies1} Yu.~E.~Lozovik and V.~I.~Yudson, Pis'ma Zh.\ Eksp.\ Teor.\
Fiz.\ {\bf 22}, 556 (1975) [JETP Lett.\ {\bf 22}, 271 (1975)]; Zh. Eksp.\ Teor.\
Fiz.\ {\bf 71}, 738 (1976) [Sov.\ Phys.\ JETP {\bf 44}, 389 (1976)]; S.~I.~Shevchenko,
Fiz.\ Nizk.\ Temp.\ {\bf 2},505 (1976) [Sov.~J.~Low Temp.~Phys. {\bf 2}, 251
(1976)].

\bibitem{exp} T.~Fukuzawa, E.~E.~Mendez, and J.~M.~Hong, Phys.~Rev.~Letts. {\bf 64}, 
3066 (1990); J.-P.~Cheng {\em et al.}, {\em ibid.} {\bf 74},450 (1995);
L.~V.~Butov and A.~I.~Filin, Phys.\ Rev.\ B {\bf 58}, 1980 (1998);

\bibitem{vign96} G.~Vignale and A.~H.~MacDonald, Phys.\ Rev.\ Lett.\ {\bf 76},
2786 (1996).

\bibitem{dragexp} See {\em e.g.}, X. G. Feng {\em et al.}, Phys.\ Rev.\ Lett.\
{\bf 81}, 3219 (1998), and references therein.

\bibitem{dragtheo} See, {\em e.g.}, the review by A.~G.~Rojo, J.\ Phys.\ Cond.\ Mat.\ 
{\bf 11}, R31 (1999), and references therein.

\bibitem{swie95} L. \'Swierkowski, J. szyma\'nsky, and Z. W. Gortel, Phys.\ Rev.\
Lett.\ {\bf 74}, 3245 (1995); Surf.\ Sci. {\bf 361/362}, 130 (1996). 
The authors treated correlations using a local field correction term, and found 
{\em inter}layer correlations had a negligible effect on the drag. 

\bibitem{uddi98} I.~Ussishkin and A.~Stern, Phys.\ Rev.\ Lett.\ {\bf 81}, 3932 
(1998); F.~Zhou and Y.~B.~Kim, Phys.\ Rev.\ B {\bf 59}, R7825 (1999).

\bibitem{lill98} M.~P.~Lilly, J.~P.~Eisenstein, L.~N.~Pfeiffer, and K.~W.~West,
Phys.\ Rev.\ Lett.\ {\bf 80}, 1714 (1998).

\bibitem{gorn99} For a treatment of Coulomb drag with correlated disorder, see
I.~V.~Gornyi, A.~G.~Yakenskin, and D.~V.~Khveschencko, Phys.\ Rev.\ Lett.\ {\bf 83},
152 (1999).

\bibitem{flen95} K.~Flensberg, B.~Y.-K.~Hu, A.~P.~Jauho, and J.~Kinaret,
Phys.\ Rev.\ B {\bf 52}, 14761 (1995).

\bibitem{kame95}  A.~Kamenev and Y.~Oreg, Phys.\ Rev.\ B
{\bf 52}, 7516 (1995).

\bibitem{crav73} R.~A.~Craven, G.~A.~Thomas, and R.~D.~Parks, Phys.\ Rev.\ 
B {\bf 7}, 157 (1973).

\bibitem{bones} N.~E.~Bonesteel, Phys.\ Rev.\ B {\bf 48}, 11484 (1993);
N.~E.~Bonesteel, I.~A.~MacDonald, and C.~Nayak, Phys.\ Rev.\ Lett.\ {\bf 77}, 
3009 (1996).

\bibitem{kell72} J.~Keller and V.~Korenman, Phys.\ Rev.\ B {\bf 5}, 4367 
(1972).

\bibitem{maha90a} See {\em e.g.}, G.~D.~Mahan, Many-Particle Physics 
(Plenum, New York, 1990), Chapter 3.

\bibitem{simplicity}
For simplicity, current vertex corrections are ignored; {\em i.e.}, 
the impurities are assumed to be $\delta$-function 
scatterers so that the transport and life times are the same.
Generalization to non-$\delta$-function impurities is not difficult
(see Ref.\ \onlinecite{flen95}).

\bibitem{maha90b} Ref.~\onlinecite{maha90a}, pg.~615.

\bibitem{kadanoff}
L.~P. Kadanoff and G.~Baym {\em Quantum Statistical Mechanics}
(Addison-Wesley, Reading, Mass., 1989).

\bibitem{jauh93}
A.-P.~Jauho and H.~Smith, Phys.\ Rev.\ B {\bf 47},  4420  (1993); 
K.~Flensberg and B.~Y.-K.~Hu, Phys.\ Rev.\ B {\bf 52}, 14796 (1995).

\bibitem{siva92}
U. Sivan, P.~M. Solomon, and H. Shtrikman, Phys.\ Rev.\ Lett. {\bf 68},  1196 (1992).

\bibitem{zhen93}
L. Zheng and A.~H. MacDonald, Phys.\ Rev.\ B {\bf 48},  8203  (1993).

\bibitem{yang99} K.~Yang and A.~H.~MacDonald, archived at cond-mat/9904061.

\bibitem{schr64} See {\em e.g.}, J. R. Schrieffer, Theory of Superconductivity
(Benjamin, Reading, 1964) pg. 164.

\bibitem{assump} Note that these were using the assumption the $\tau^{-1} < T$.
For $\tau^{-1} > T$, $\tau^{-1}$ replaces $T$,
the numerical values of the coefficients change, and one would
regain the diffusion result for the Cooperon.

\bibitem{cont98} S.~Conti, G.~Vignale, and A.~H.~MacDonald, Phys.\ Rev.\ 
B {\bf 57}, 6846 (1998).

\bibitem{bons98} M.~C.~B\o nsager {\em et al.}, Phys.\ Rev.\ B {\bf 57},
7085 (1998).

\end{thebibliography}
\end{document}